\documentclass[11pt,dvips]{article}
\usepackage{epsfig,times}
\usepackage{picinpar}
\usepackage{wrapfig}
\usepackage{floatflt}

\makeatletter
\renewcommand{\section}{\@startsection{section}{1}{0in}
        {0.4\baselineskip}{0.1\baselineskip}{\Large\bf}}
\renewcommand{\subsection}{\@startsection{subsection}{2}{0in}
        {0.25\baselineskip}{-\baselineskip}{\large\bf}}
\renewcommand{\subsubsection}{\@startsection{subsubsection}{3}{0in}
        {0.1\baselineskip}{-\baselineskip}{\normalsize\bf}}
\makeatother

\def\beq{\begin{equation}}
\def\eeq{\end{equation}}
\def\ni{\noindent}

\begin{document}

\makeatletter\newcommand{\ps@icrc}{
\renewcommand{\@oddhead}{\slshape{HE.6.1.17}\hfil}}
\makeatother\thispagestyle{icrc}
\markright{HE.6.1.17}

\begin{center}
{\LARGE \bf  Performance of RPCs operated at the Yangbajing Laboratory}
\end{center}

\begin{center}
{\bf ARGO-YBJ Collaboration}, presented by S.M. Mari$^{1}$\\
{\it $^{1}$ Universit\'a della Basilicata, Potenza (Italy)}
\end{center}

\begin{center}
{\large \bf Abstract\\}
\end{center}
\vspace{-0.5ex}

The ARGO-YBJ experiment will be installed at YangBaJing Cosmic Ray 
Laboratory (Tibet, China), 4300 m a.s.l . It consists of a full coverage of 
$\sim 10^4$ $m^2$ realized with RPC chambers. A small carpet of 
$\sim 50$ $m^2$ has been operated at YangBaJing in order to check the RPC 
performance in these high altitude conditions. 
Results concerning efficiency and time resolution are reported.
\vspace{1ex}

\section{Introduction}

The aim of the ARGO-YBJ experiment (Abbrescia et al. (1996)) is the study of 
cosmic rays, mainly $\gamma$-radiation, in an energy range down to 
$\sim 100$ $GeV$, by detecting small size air showers with a ground detector.
This very low energy threshold, which is below the upper limit of
the next generation satellite experiments, is achieved in two ways: 
\begin{verse}
1) By operating the experiment at very high altitude to better
approach the level where low energy air showers reach their
maximum development. The choice of the YangBaJing (YBJ) Cosmic Ray
Laboratory (Tibet, China), 4300 m a.s.l, was found to be very appropriate.\\
2) By utilizing a full coverage detector to maximize the number of detected
particles for a small size shower.
\end{verse}

The ARGO-YBJ detector consists of a single RPC layer of $\sim
5000$ $m^2$ and $\sim 92 \%$ coverage, surrounded by a ring of
sampling stations which recover edge effects and increase the
sampling area for showers initiated by $>5$ $TeV$ primaries.

The trigger and the DAQ systems are built following a two level architecture. 
The signals of a set of 12 contiguous RPCs, referred to as a Cluster in the
following, are managed by a Local Station.
The information from each Local Station is collected and elaborated in the
Central Station. According to this logic a Cluster represents the basic 
detection unit.

A Cluster prototype of 15 RPCs has been put in operation in the YBJ 
Laboratory in order to check both the performance of RPCs operated in a high 
altitude site and their capability of imaging a small portion of the shower 
front.

In this paper the results concerning the performance of $2$ $mm$ gap, 
bakelite RPC
detectors operated in streamer mode at an atmospheric depth of $606$ $g/cm^2$
are described. Data collected by the carpet and results from their analysis
are presented elsewhere (D'Ettorre Piazzoli et al. (1999)).

\section{The experimental set-up}

The detector, consisting of a single gap RPC layer, is installed
inside a dedicated building at the YBJ Laboratory. The set-up 
is an array of 3x5 chambers of area $280 \times 112$ $cm^{2}$ each, 
laying on the building floor and covering a
total area of $8.7 \times 6.1$ $m^{2}$. The active area of $46.2$ 
$m^{2}$, accounting for a dead area due to a $7$ $mm$ frame closing
the chamber edge, corresponds to a $90.6 \%$ coverage. The RPCs,
with a $2$ $mm$ gas gap, are built with bakelite electrode plates of
volume resistivity in the  range ($0.5\div 1$)  $ 10^{12} \Omega
\cdot cm$, according to the standard scheme reported in (ATLAS (1997)).

The RPC signals are picked up by means of aluminum strips $3.3$ $cm$
wide and $56$ $cm$ long which are glued on a $0.2$ $mm$ thick film of
Poly-Ethylen-Tereftalate (PET). The strips are embodied in a panel,
consisting of a $4$ $mm$ thick polystyrene foam sheet sandwiched
between the PET film and an aluminum foil used as a ground
reference electrode. At the edge of the
detector the strips are connected to the front end electronics and
terminated with 50 $\Omega$ resistors. A grounded aluminum foil is used to
shield the bottom face of the RPC and an extra PET foil, on top of
the aluminum, is used as a further high voltage insulator. 
The front end circuit contains 16 discriminators, with about $50$ $mV$ 
voltage threshold, and provides a FAST-OR signal with the same
input-to-output delay ($10$ $ns$) for all the channels. This signal is used
for time measurements and trigger purposes in the present test.
The 16 strips connected to the same front end board are logically
organized in a pad of $56 \times 56$ $cm^{2}$ area. Each RPC is
therefore subdivided in 10 pads which work like independent
functional units. The pads are the basic elements which define the
space-time pattern of the shower; they give indeed the position
and the time of each detected hit. The FAST-OR signals of all 150
pads are sent through coaxial cables of the same length to the
carpet central trigger and read out electronics.
The trigger logics allows to select events with a pad multiplicity
in excess of a given threshold. At any trigger occurrence the
times of all the pads are read out by means of multihit TDCs of $1$ 
$ns$ time bin, operated in common STOP mode. 
The set-up was completed with a small telescope consisting of 3 RPCs of 
$50 \times 50$ $cm^{2}$ area with 16 pick-up 
strips $3$ $cm$ wide connected to front end electronics board
similar to the ones used in the carpet. The 3 RPCs were overlapped
one on the other and the triple coincidence of their FAST-OR
signals was used to define a cosmic ray crossing the telescope.

\section {Experimental results}

The measurements described in this paper were performed in the
second half of February 1998 with an external temperature in the 
range $-20\div -5$ $C$ and in the first half of May when the
temperature was in the range $-5\div +15$ $C$. The internal temperature
was kept, by using some heaters, between $\sim +4\div +8$ $C$ in the first
run and $\sim 16\div 18$ $C$ in the second. The laboratory temperature
and pressure were monitored during all data taking.
The RPCs of the test carpet were operated in streamer mode 
as foreseen for the final experiment. This mode delivers
large amplitude saturated signals, and is less sensitive than
the avalanche or proportional mode to electromagnetic
noise, to changes in the environment conditions and to mechanical
deformations of the detector. On the other hand the larger rate
capability achievable in avalanche mode is not needed in
a cosmic ray experiment.

Three gas mixtures were tested which used the same components, Argon, 
Isobutane and Tetrafluoroethane, in different proportions:
TFE/Ar/i-But = 45/45/10; 60/17/13 and 75/15/10. In the three cases
the ratio Ar/TFE was changed to a large extent, living the i-But
concentration relatively stable.
TFE is an heavy gas with good quenching properties (Cardarelli et al. (1996)). 
An increase of TFE concentration in place of the Ar concentration
should therefore increase the primary ionization thus compensating
for the $40\%$ reduction caused by the lower gas target pressure
(600 $mbar$) and reduce the afterpulse probability. For each of the
three gases a voltage scan was made for RPC2 (the RPC in the central position 
of the telescope), leaving the other
two RPCs at a fixed operating voltage, and the following
measurements were made: RPC2 counting rate, current and efficiency. 
The detection efficiency $vs$
the operating voltage for the three gases is shown in Fig. 1. 
The reduction of the Argon concentration in favor of TFE results
in a clear increase of the operating voltage as expected from the
large quenching action of TFE.

\begin{figure}[htb]
\vfill \begin{minipage}{.47\linewidth}
\begin{center}
\mbox{\epsfig{file=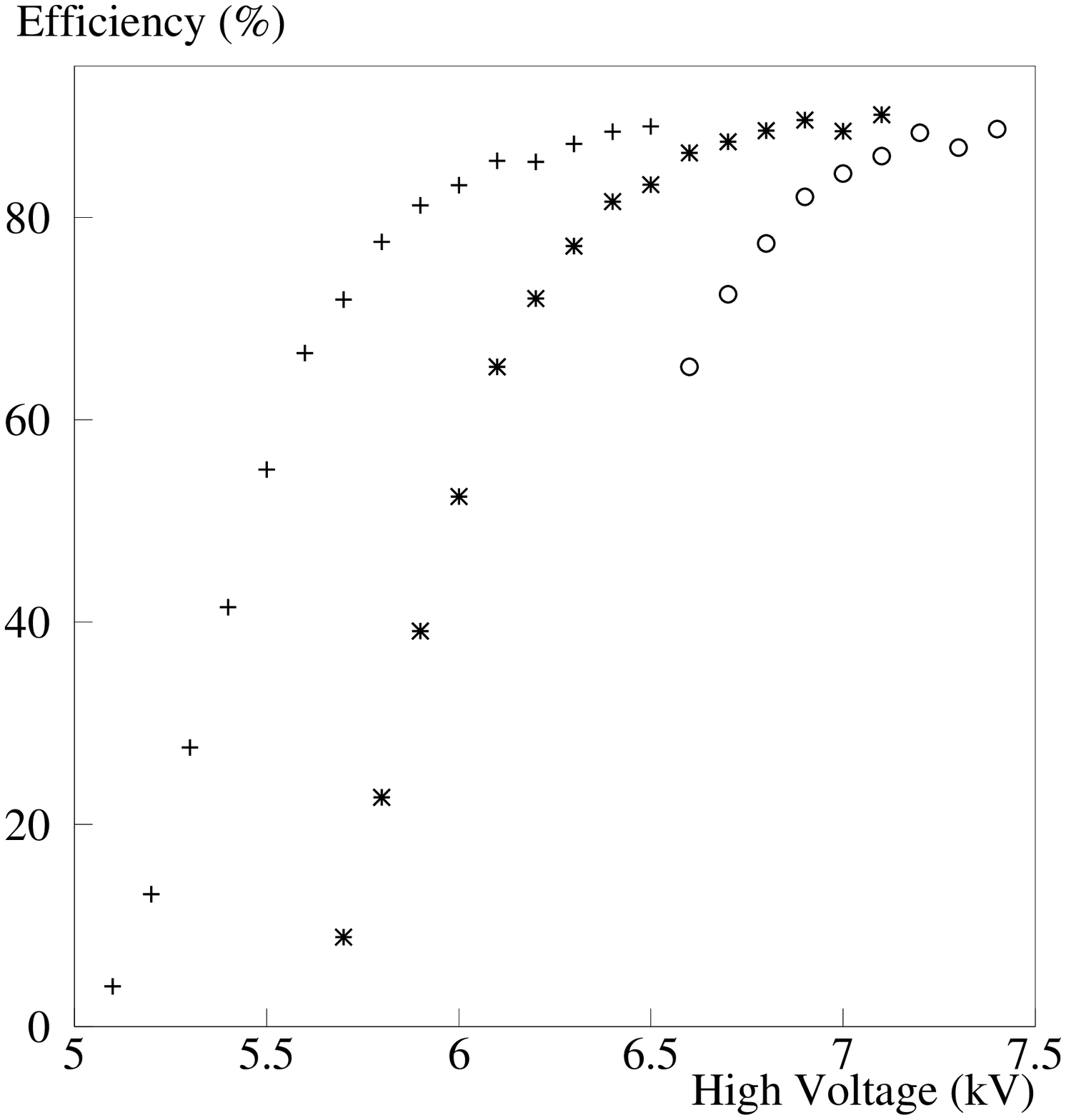,height=8.cm,width=8.cm}}
\end{center}
\caption{\em Detection efficiency of one RPC of the auxiliary
    telescope vs operating voltage for 3 gases: TFE/Ar/i-But=45/45/10
    (+); 60/17/13 (*) and 75/15/10 ($\circ$)}
\end{minipage}\hfill
\hspace{-0.5cm}
\begin{minipage}{.47\linewidth}
\begin{center}
\mbox{\epsfig{file=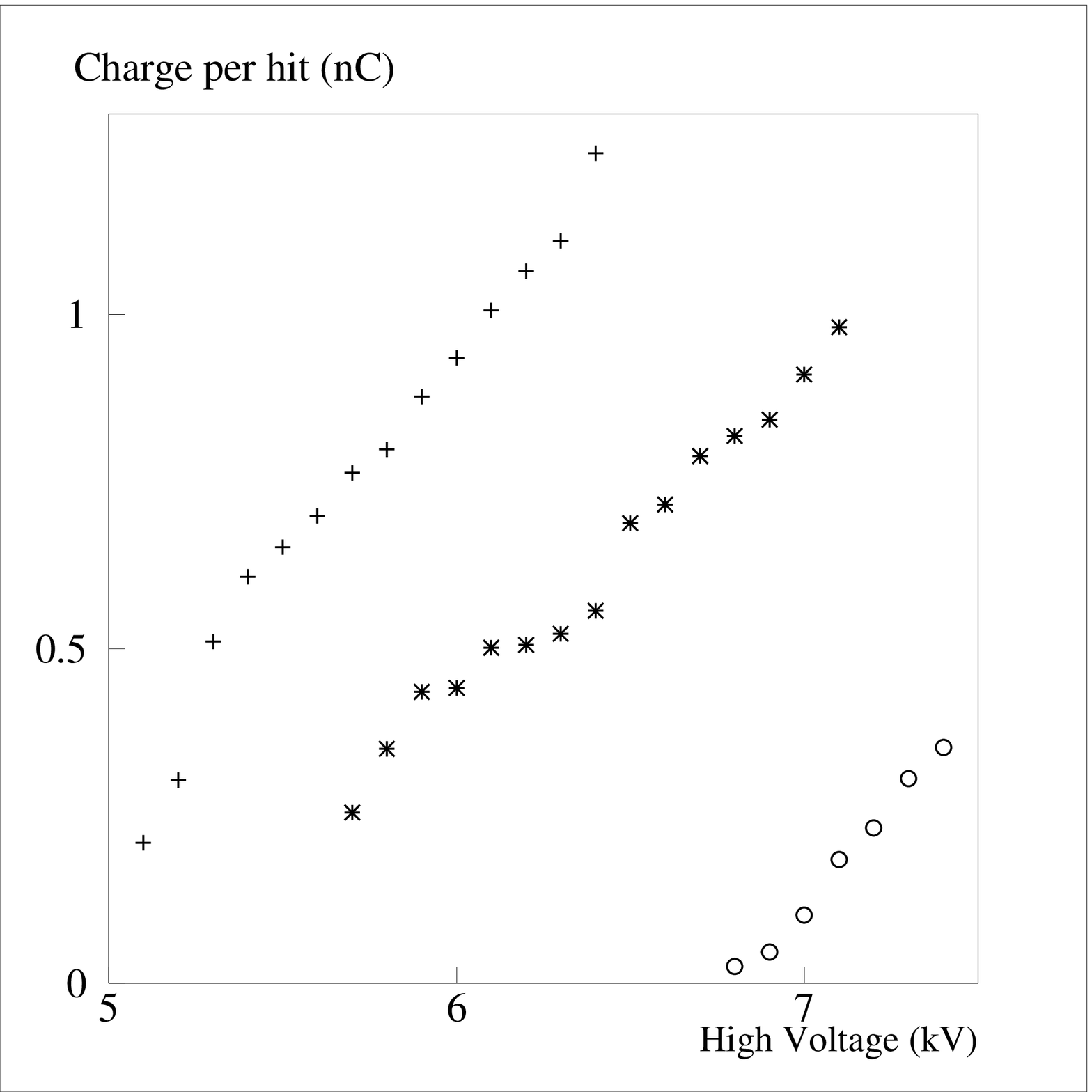,height=8.cm,width=8.cm}}
\end{center}
\caption{\em Charge delivered per count for the 3 gases vs operating 
voltage: TFE/Ar/i-But=45/45/10
    (+); 60/17/13 (*) and 75/15/10 ($\circ$)}
\end{minipage}\hfill
\end{figure}

In spite of the different operating voltages all three gases approach
the same efficiency of $\sim 90\%$ which include the inefficiency
due to geometrical effects. The ratio of the operating current to the
counting rate gives the charge per count delivered in the RPC gas,
which is shown in Fig. 2 as a function of the operating
voltage for the three gases. Here a small term corresponding to
the current leaks is subtracted to the total current. 
The data shows that the higher the TFE fraction, the lower is the charge 
delivered in the gas by a single streamer. Since a lower delivered charge 
extends the dynamic range achievable for the analogical read-out, we decided 
to operate the test carpet with the gas mixture corresponding to the 
highest TFE fraction.

\begin{figure}[htb]
\vfill \begin{minipage}{.47\linewidth}
\begin{center}
\mbox{\epsfig{file=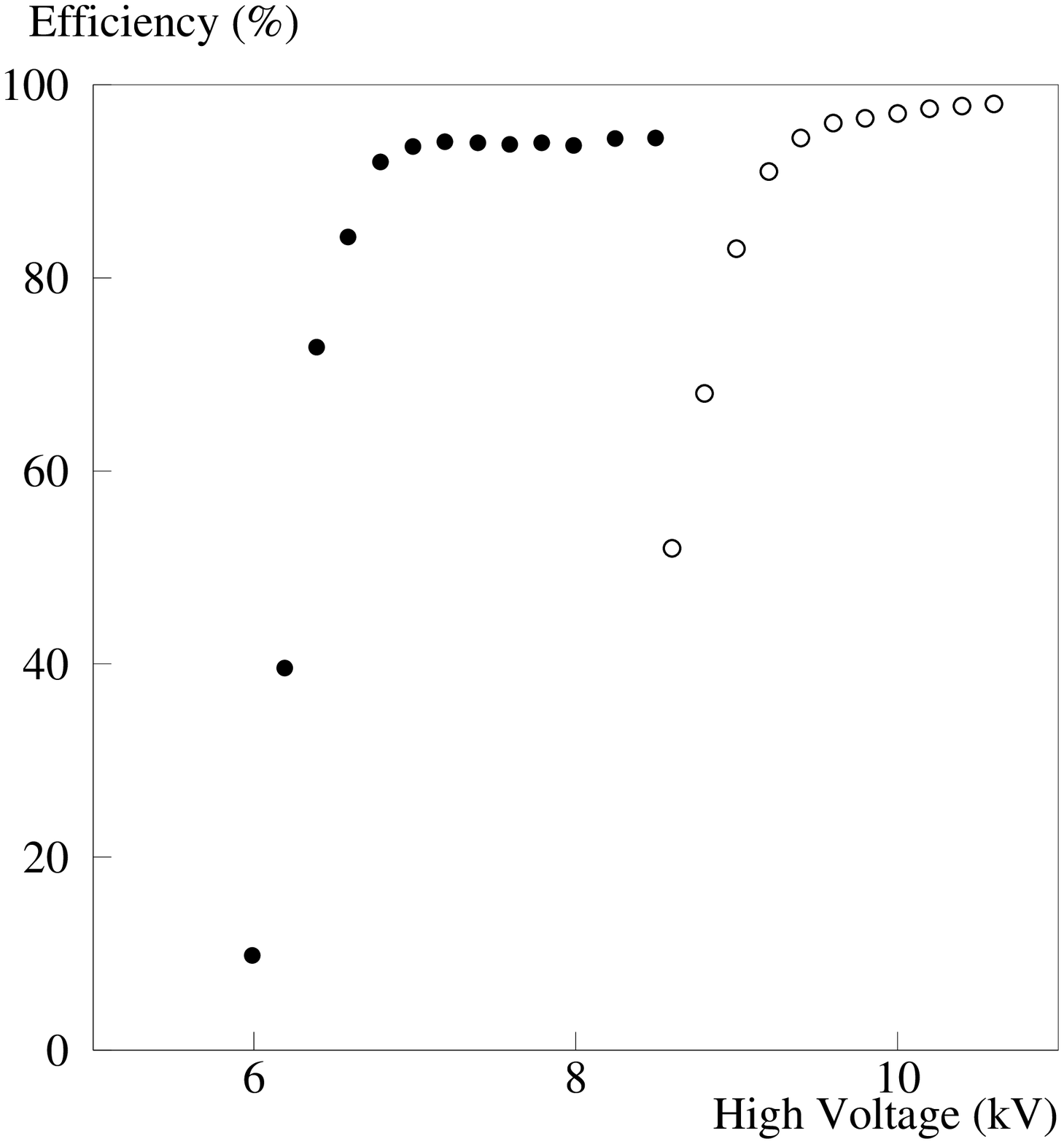,height=8.cm,width=8.cm}}
\end{center}
\caption{\em Detection efficiency vs operating voltage for one of the 
carpet RPCs ($\bullet$). The same curve for a 2 mm gap RPC operating at
sea level is also reported ($\circ$) for comparison.}
\end{minipage}\hfill
\hspace{-0.5cm}
\begin{minipage}{.47\linewidth}
\begin{center}
\mbox{\epsfig{file=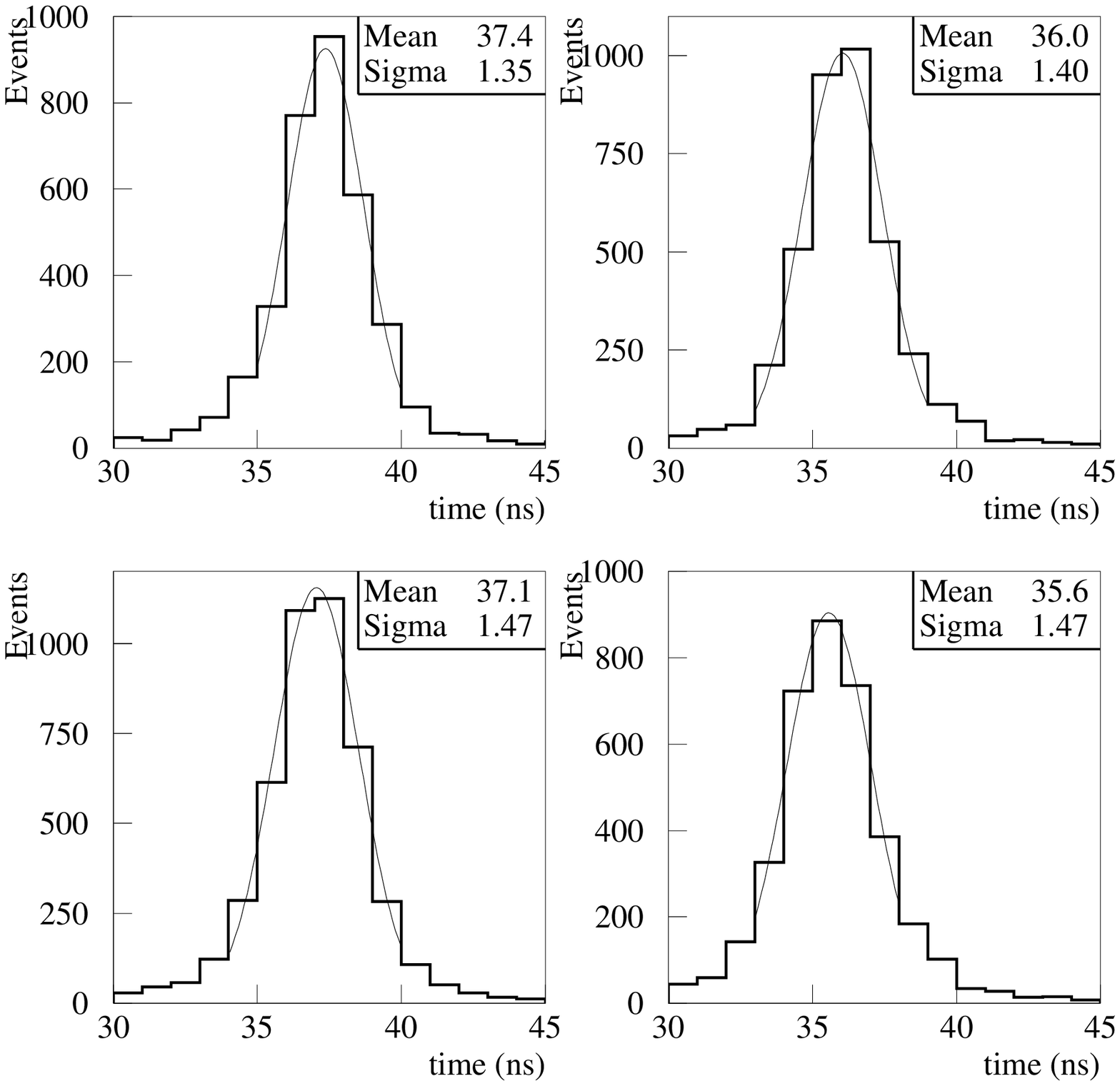,height=8.cm,width=8.cm}}
\end{center}
    \caption{\em Time jitter distribution of 4 pads of the carpet.
The telescope RPC2 signal is used as Common Stop. }
\end{minipage}\hfill
\end{figure}

Fig. 3 shows the operating
efficiency for the ORed pads 2-3-7-8 of one RPC of the carpet. The
efficiency was measured using cosmic ray signals defined by the
triple coincidence of the RPCs of the auxiliary telescope which
was placed on top of the carpet and centered on the corner among
the pads 2-3-7-8. The same curve for a $2$ $mm$ gap RPC operated at sea level 
is also shown for comparison. The detection efficiency vs operating voltage, 
compared to the operation at $606$ $mbar$ pressure in YBJ, 
shows an increase of $\sim 2.5$ $kV$ in operating voltage. The effect
of small changes in temperature $T$ and pressure $P$ on the operating
voltage can be accounted for by rescaling the applied
voltage $V_a$ according to the relationship:

\beq
  V=V_{a}\frac{P_{0}}{P}\cdot \frac{T}{T_{0}}
\eeq

\ni where $P_0$ and $T_0$ are arbitrary standard values, e.g. $1010$ $mbar$ and
$293$ $K$ respectively for a sea level laboratory. This formula predicts,
starting from the YBJ data, an operating voltage at sea level
which is considerably smaller than the experimental one.
However, a good consistency is recovered by assuming that, in the ideal gas 
approximation, the parameter which fixes the operating
voltage is given by {\em gap $\cdot$ pressure / temperature}.
Measurements on RPCs with different gas gap size justify this 
assumption (Bacci et al. (1999)). 

Fig. 3 also shows that the plateau efficiency measured at
YBJ is $3-4\%$ lower than at the sea level. Although a lower
efficiency is expected from the smaller number of primary clusters
at the YBJ pressure, we attribute most of the difference to the underestimation 
of the YBJ efficiency. At the YBJ level
indeed the ratio of the cosmic radiation electromagnetic to muon
component is $\sim 4$ times larger than at sea level. A spatial tracking
with redefinition of the track downstream of the carpet would
eliminate the contamination from soft particles, giving  a more
accurate and higher efficiency. On the other hand the lower
efficiency could hardly be explained with the gas lower density.
The number of primary clusters in the YBJ test, estimated around
$9$, is the same as in the case of some gas, e.g.
Ar/i-But/CF3Br = 60/37/3, that was frequently used at sea level with
efficiency of $\sim 97-98\%$.

A rather flat singles counting rate plateau is observed at a level 
of $\sim 400$ $Hz$ for a single pad of area 56$\times$56 $cm^{2}$. 

The time jitter distribution of the
pad signals was obtained by measuring the delay of the FAST-OR signal
with respect to RPC2 in the trigger telescope by means of a TDC
with 1 $ns$ clock.  This distribution is shown in Fig. 4 for
the four pads. The average of the standard deviations is $1.42$ $ns$
corresponding to a resolution of $\sim 1$ $ns$ for the single RPC if we
account for the fact that the distributions show the combined jitter of two 
detectors.

\section  {Summary}

The use of RPCs in high
altitude laboratories poses some basic questions concerning how 
the operating voltage, the plateau efficiency and the time resolution do scale 
with the pressure for the streamer mode operation.

Data collected at the Yangbajing Laboratory with a RPC carpet of 
$\sim 50$ $m^2$ and with a small RPC telescope of area $50\times 50$ $cm^2$ 
confirm that this detector can be operated efficiently ($\geq 95\%$) 
at high altitude with excellent time resolution ($\sim 1$ $ns$). The shift 
of the operating point is well understood. Other noticeable advantages as 
low cost, large active area, pixel size defined by external electrodes and 
the possibility of an easy integration in large systems, make this detector 
well suitable for using in EAS physics. 

\vspace{1ex}
\begin{center}
{\Large\bf References} 
\end{center}
Abbrescia M. et al., {\it Astroparticle Physics with ARGO}, Proposal (1996).
This document can be downloaded at the URL:
http://www1.na.infn.it/wsubnucl/cosm/argo/argo.html\\
D'Ettorre Piazzoli B. et al. (ARGO-YBJ coll.), (1999) these proceed. HE.6.1.03\\
ATLAS Muon Spectrometer Technical Design Report, (Chapter 8), CERN/LHCC/97-22.\\
R. Cardarelli, V. Makeev, R. Santonico, Nucl. Instr. Meth. {\bf A382} (1996) 
470.\\
Bacci C. et al. (ARGO-YBJ coll.), (1999) submitted to Nucl. Instr. Meth.\\
\end{document}